\documentclass[aps,prc,groupedaddress,showpacs,manuscript]{revtex4}

\usepackage{graphicx}

\begin{document}

\title{Density and temperature dependence of nucleon-nucleon elastic cross section}

\author {Qingfeng LI$^{1)}$
\email[]{liqf@itp.ac.cn}, Zhuxia LI$^{2,1,3)}$
\email[]{lizwux@iris.ciae.ac.cn}, and Enguang ZHAO$^{1)}$}
\address{
1) Institute of Theoretical Physics,
Chinese Academy of Sciences, P. O. Box 2735, Beijing 100080, P. R. China\\
2) China Institute of Atomic Energy, P. O. Box 275 (18),
Beijing 102413, P. R. China\\
3) Center of Theoretical Nuclear Physics, National Laboratory of
Lanzhou Heavy Ion Accelerator,
 Lanzhou 730000, P. R. China
}


\begin{abstract}
The in-medium neutron-proton, proton-proton (neutron-neutron)
elastic scattering cross sections ($\sigma_{np}^{*}$,
$\sigma_{pp(nn)}^{*}$) are studied based on the effective
Lagrangian of density dependent relativistic hadron theory in
which the $\delta$[$a_0(980)$] meson is included. Our study shows
that at low densities the $\sigma_{np}^*$ is about $3-4$ times
larger than $\sigma_{pp(nn)}^*$ and at densities higher than the
normal density the isospin effect is almost washed out. Because of
coupling to $\delta$ meson the $\sigma_{nn}^*$ and $\sigma_{pp}^*$
are different in isospin asymmetric medium following the splitting
of the proton and neutron mass. The isospin effect on the density
dependence of the in-medium nucleon elastic cross section is
dominantly contributed by the isovector $\delta$ and $\rho$
mesons. The temperature effect on the $\sigma_{np}^*$ and
$\sigma_{pp(nn)}^*$ is studied. It is shown that the temperature
effect is weaker compared with the density effect but it becomes
obvious as density increases.
\end{abstract}


\maketitle

The rapid advance in nuclear reactions using rare isotopes has
opened up several new frontiers in nuclear science
\cite{LiB01,Pa97,Ye01,Ts01,Di01,So03}. In particular, the
intermediate energy heavy rare isotopes currently available at
National Superconducting Cyclotron Laboratory (NSCL/MSU) and the
future new facilities RIA in USA and SIS-200 at GSI in Germany
provide a unique opportunity to explore novel properties of dense
neutron-rich matter that was not in reach in terrestrial
laboratories before.

In order to study theoretically the neutron-rich nuclear
collisions at intermediate energy within a microscopic transport
model approach, both the isospin-dependent mean field and the
two-body scattering cross sections should be introduced. The
isospin dependent mean field has been studied extensively with
both non-relativistic and relativistic theory. Concerning the
nucleon-nucleon elastic cross sections (ECS), it is already known
that up to hundreds MeV the free proton-neutron cross section is
about $2-3$ time larger than that of proton-proton
(neutron-neutron)'s \cite{Ch68}. However, it is not clear how the
nuclear medium corrects the neutron-proton and proton-proton (or
neutron-neutron) ECS. The isospin dependent in-medium ECS were
calculated based on QHD-II model in which the isospin-vector
channel was introduced through a coupling to the vector-isovector
$\rho$ meson field \cite{Li00}. In this work we intend to
introduce the scalar-isovector $\delta[a_0(980)]$ meson into the
effective Lagrangian to calculate the in-medium two-body
scattering elastic cross sections. In free NN potentials the
$\delta$ meson is a standard and essential ingredient. The
importance of the $\delta$ meson in nuclear matter with extreme
neutron-to-proton ratios was conjectured in
\cite{Hu96,Kubis97,Hu98}. It was pointed out that conventional
mean-field models, neglecting the $\delta$ meson, are likely to
miss an important contribution to the isospin degree of freedom
which ought to be necessary for a proper description of strongly
asymmetric matter. Recently, the role of the $\delta$ meson for
asymmetric nuclear matter was studied by \cite{Liu02}. Some
calculation results for the structure of exotic nuclei showed the
importance of the inclusion of the $\delta$ meson for the
stability conditions of drip-line nuclei \cite{Hofmann01, Leja01}.
Therefore it would be worthwhile to study how the inclusion of the
$\delta$ meson changes the medium effect on nucleon-nucleon cross
sections. We take the  same approach as in
\cite{Mao94,Mao942,Li00} for the formalism and the effective
lagrangian employed in the density dependent relativistic hadron
theory for asymmetric nuclear matter and exotic nuclei given in
\cite{Hofmann01}.

The effective Lagrangian density is taken as
\begin{eqnarray}
L^{} &=&\bar{\Psi}[i\gamma _\mu \partial ^\mu -M_N]\Psi +\frac
12\partial _\mu \sigma \partial ^\mu \sigma -\frac 14F_{\mu \nu
}\cdot F^{\mu \nu }
\nonumber \\
&&+\frac {1}{2} \partial _{\mu} \mbox{\boldmath $\delta$}\partial
^{\mu} \mbox {\boldmath $\delta$}-\frac 14L_{\mu \nu }\cdot L^{\mu
\nu } -\frac 12m_\sigma ^2\sigma ^2+\frac 12m_\omega ^2\omega _\mu
\omega ^\mu -\frac 12m_\delta ^2\mbox{\boldmath $\delta$}^2 +
\frac 12m_\rho ^2\mbox{\boldmath $\rho$}_\mu \mbox{\boldmath
$\rho$ }^\mu \nonumber \\
   &&+g_\sigma \bar{\Psi}\Psi
\sigma -g_\omega \bar{\Psi}\gamma _\mu \Psi \omega ^\mu  \nonumber
+g_\delta \bar{\Psi}\mbox{\boldmath $\tau$}\cdot \Psi
 \mbox{\boldmath $\delta$}-\frac 12g_\rho \bar{\Psi}\gamma _\mu %
\mbox{\boldmath $\tau$}\cdot \Psi \mbox{\boldmath $\rho$}^\mu,
\end{eqnarray}
where $g_{\sigma}$, $g_{\omega}$, $g_{\rho}$, and $g_{\delta}$ are
density dependent and $F_{\mu \nu }\equiv \partial _\mu \omega
_\nu -\partial _\nu \omega _\mu$, $L_{\mu \nu }\equiv \partial _\mu \mbox{\boldmath $\rho$}_\nu -\partial _\nu %
\mbox{\boldmath $\rho$}_\mu $.

The distribution functions for fermions ($f({\bf p})$) and
antifermions ($\bar{f}({\bf p})$) respectively are $f_i({\bf
p})=1/(1+e^{(E_i^*({\bf p})-\mu_i^*)/T})$, $\bar{f}_i({\bf
p})=1/(1+e^{(E_i^*({\bf p})+\mu_i^*)/T})$, where $i$ represents
proton ($i=p$) or neutron ($i=n$) and $E^*=\sqrt{{\bf
p}^2+{M_i^*}^2}$. The effective nucleon mass $M^{*}$ is given by
\begin{equation}
M^*=M_0+\Sigma_{H(\sigma) }(x,\tau)+\Sigma_{H(\delta ^0)}(
x,\tau). \label{eqms}
\end{equation}
$\Sigma_{H(\sigma) }$ and $\Sigma_{H(\delta ) }$ are the
self-energy parts of nucleon contributed from $\sigma$ and
$\delta$ meson, respectively. Because the self-energy
$\Sigma_{H(\delta ^0)}( x,\tau)$ has opposite sign for neutron and
proton for isospin asymmetric medium the correction of the nuclear
medium to proton mass and neutron mass from $\delta$ meson is of
opposite sign. Thus the proton and neutron effective mass differ
for isospin asymmetric systems. The effective chemical potential
$\mu^*$ is
\begin{equation}
\mu_i^*=\mu_i+\Sigma_{H(\omega)}^0(x,\tau)+\Sigma_{H(\rho^0)}^0(x,\tau).
\label{mui}
\end{equation}
$\Sigma_{H(\omega) }$ and $\Sigma_{H(\rho) }$ are the self-energy
parts of nucleon contributed from $\omega$ and $\rho$ meson,
respectively.
 Within the mean field approximation the equation of state can be written as
\begin{eqnarray}
\epsilon&=&\sum\limits_{i=n,p}2\int{\frac{d^3{\bf
p}}{(2\pi)^3}}E_i^*({\bf p})[f_i({\bf p})+\bar{f}_i({\bf p})]+
U(\sigma)\nonumber\\
&&+\frac12\Sigma_{H(\omega)}^0(x,\tau)\rho_B+\frac12\Sigma_{H(\rho^0)}^0(x,\tau)\rho_{B3}-
\frac12\Sigma_{H(\delta^0)}(x,\tau)\rho_{S3} ,
\label{epsilon}
\end{eqnarray}
where the baryon and scalar densities $\rho_B$ and $\rho_S$ are
\begin{equation}
\rho_B=\sum\limits_{i=n,p}2\int{\frac{d^3{\bf
p}}{(2\pi)^3}}[f_i({\bf p})-\bar{f}_i({\bf p})], \label{rhob}
\end{equation}
\begin{equation}
\rho_S=\sum\limits_{i=n,p}2\int{\frac{d^3{\bf
p}}{(2\pi)^3}}\frac{M_i^*}{E_i^*}[f_i({\bf p})+\bar{f}_i({\bf p})]
\label{rhos}
\end{equation}
while ${\rho_B}_3={\rho_B}_p-{\rho_B}_n$ and ${\rho_S}_3={\rho_S}_p-{\rho_S}_n$.

The functional form of the density dependent coupling for
$\sigma$, $\omega$, $\delta$, and $\rho$ meson taken from Ref.
\cite{Hofmann01} is
$g_\alpha(\rho)=A_\alpha(1+B_\alpha(\rho/\rho_0+D_\alpha)^2)/(1+C_\alpha(\rho/\rho_0+E_\alpha)^2)
$, where $\alpha$ represents $\sigma$, $\omega$, $\delta$, or
$\rho$, and the parameters $A_\alpha$, $B_\alpha$, $C_\alpha$,
$D_\alpha$, and $E_\alpha$ are listed in Table \ref{tab2}.

\begin{table}

\caption{Parametrization of the density dependent couplings taken
from Ref. \cite{Hofmann01}.}

\begin{tabular}{cllll}
\hline\hline
 Meson $\alpha$ & \hspace{1.cm}$\sigma$ \hspace{1.cm} & \hspace{1.cm}$\omega$\hspace{1.cm} &
\hspace{1.cm}$\delta$\hspace{1.cm} &
\hspace{1.cm}$\rho$\hspace{1.cm}
\\ \hline
$m_\alpha$[MeV]&\hspace{1.cm}550\hspace{1.cm}&\hspace{1.cm}783\hspace{1.cm}&\hspace{1.cm}983\hspace{1.cm}&
\hspace{1.cm}770\hspace{1.cm}\\
\hline $A_\alpha$ & \hspace{1.cm}13.1334\hspace{1.cm} &
\hspace{1.cm}15.1640\hspace{1.cm} &
\hspace{1.cm}19.1023\hspace{1.cm} & \hspace{1.cm}19.6270\hspace{1.cm} \\
$B_\alpha$ & \hspace{1.cm}0.4258\hspace{1.cm} & \hspace{1.cm}0.3474\hspace{1.cm} & \hspace{1.cm}1.3653\hspace{1.cm} & \hspace{1.cm}1.7566\hspace{1.cm} \\
$C_\alpha$ & \hspace{1.cm}0.6578\hspace{1.cm} & \hspace{1.cm}0.5152\hspace{1.cm} & \hspace{1.cm}2.3054\hspace{1.cm} & \hspace{1.cm}8.5541\hspace{1.cm} \\
$D_\alpha$ & \hspace{1.cm}0.7914\hspace{1.cm} & \hspace{1.cm}0.5989\hspace{1.cm} & \hspace{1.cm}0.0693\hspace{1.cm} & \hspace{1.cm}0.7783\hspace{1.cm} \\
$E_\alpha$ & \hspace{1.cm}0.7914\hspace{1.cm} &
\hspace{1.cm}0.5989\hspace{1.cm} & \hspace{1.cm}0.5388\hspace{1.cm} & \hspace{1.cm}0.5746\hspace{1.cm} \\
\hline
\multicolumn{5}{c}{$\rho_0: 0.18$ fm$^{-3}$; $E/A: 16$ MeV; K: 282 MeV; $M^*/M_0$: 0.554; $a_{Sym}: 26.1$ MeV.}\\
\hline\hline
\end{tabular}
\label{tab2}
\end{table}

\begin{figure}
\includegraphics[angle=0,width=0.8\textwidth]{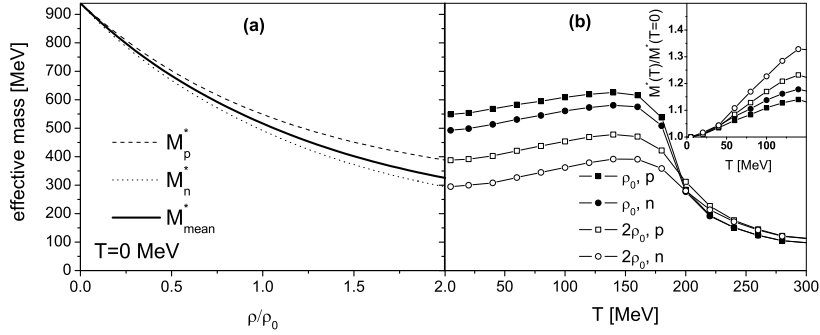}
 \caption{(a): The nucleon effective mass as a function of reduced nuclear density at $T=0$ MeV.
The effective masses for $Y_p=0.5$ is plotted as thick line, while
the proton and neutron effective masses for $Y_p=0.3$ are
demonstrated as light ones. (b): The neutron effective mass (in
circles) and proton effective mass (in rectangles) as a function
of temperature for $Y_p=0.3$. The lines with solid symbols are for
the case when $\rho/\rho_0=1$, while those with open symbols are
for the case when $\rho/\rho_0=2$. The ratios $M^*(T)/M^*(T=0)$
for different cases are shown in the upper-right plot. }
\label{fig1}
\end{figure}

By solving Eq. \ref{eqms} self-consistently we obtain the
effective proton and neutron mass. We illustrate the density
dependence of the neutron and proton effective mass at $T=0$ MeV
for $Y_p=0.3$, and $Y_p=0.5$ ($Y_p=\rho_{p}/(\rho_{p}+\rho_{n})$)
in Fig. \ref{fig1} (a). The proton and neutron effective mass
differ at $Y_p=0.3$ case due to the inclusion of the $\delta$
meson. The proton effective mass is larger than the neutron
effective mass in neutron-rich nuclear medium. With the increase
of density, the splitting of the proton and neutron mass
increases, which will affect the in-medium ECS.

Fig. \ref{fig1} (b) shows the temperature dependence of the
neutron and proton effective mass for $Y_p=0.3$. The ratios
$M^*(T)/M^*(T=0)$ for different cases are also shown in the
upper-right plot. One can see from this figure that the effective
nuclear mass first increases slightly until $T=150$ MeV and then
decreases quickly as temperature increases further, which has been
also observed in Refs. \cite{Mi02,Li96}. The sub-figure shows that
increasing slope of $M^*(T)/M^*(T=0)$ depends on density as well
as the species of nucleons. The behavior of the density and
temperature dependence of the effective mass should influence the
density and temperature dependence of in-medium ECS which will be
discussed later.

With the same approach as in \cite{Mao94,Mao942,Li00} we obtain
the expressions of $\sigma_{pp(nn)}$ and  $\sigma_{pp(nn)}$ with
inclusion of the individual $\delta$ term and the crossing terms
of $\delta$-$\sigma$ $\delta$-$\omega$ as well as $\delta$-$\rho$.
Because of the coupling to $\delta$ meson the collisions of
$np\rightarrow np$ and $np\rightarrow pn$ should be treated
differently, which makes the calculations of in-medium ECS much
more complicated. And for the same reason, $\sigma_{nn}^*$ is not
equal to  $\sigma_{pp}^*$ for isospin asymmetric systems.

Before calculating the ECS, we have to introduce the common used
form factor of nucleon-meson-nucleon vertex $F_{i}(t)=\Lambda
_i^2/(\Lambda _i^2-t)$, where the subscripts $i$ denote the
vertices for different meson-nucleon couplings and $\Lambda_{i}$
is the cut-off mass for $i$ meson species. In this work, the
values of $\Lambda _\sigma =1200 $ MeV, $\Lambda _\omega = 808 $
MeV, (which are same as those chosen in \cite{BASS98}), $\Lambda
_\delta =1000$ MeV and $\Lambda _\rho =800$ MeV are used. We find
there is very week influence on the results when taking $\Lambda
_\delta=800-1000$MeV and $\Lambda _\rho =600-800$ MeV. Again, the
$F_{i}(t)$ is different for $np\rightarrow np$ and $np\rightarrow
pn$ channels when $M_n^* \ne M_p^*$.

\begin{figure}
\includegraphics[angle=0,width=0.5\textwidth]{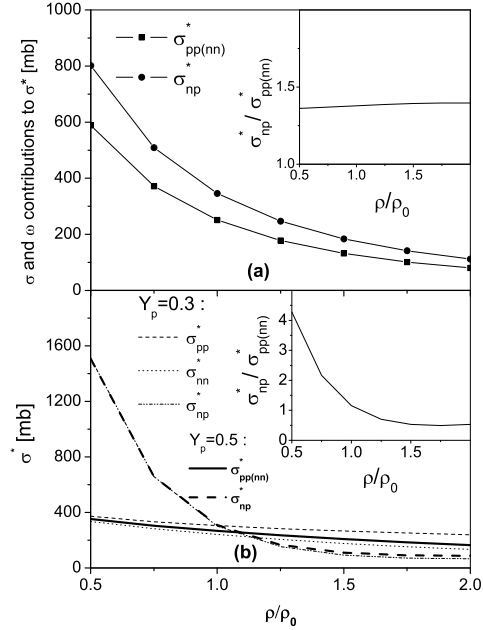}
 \caption{(a):
The $\sigma$ and $\omega$ contributions to $\sigma_{pp(nn)}$ and
$\sigma_{np}$ at $T=0$ MeV, $E_{K}=10$ MeV for $Y_p=0.5$. The
ratio between $\sigma_{np}$ and $\sigma_{pp(nn)}$ is illustrated
in the upper-right plot. (b): The density dependence of
$\sigma_{pp}^*$, $\sigma_{nn}^*$, and $\sigma_{np}^*$ in
isospin-symmetric ($Y_p=0.5$, thick lines) and asymmetric
($Y_p=0.3$, light lines) nuclear medium. $T=0$ MeV and $E_K=10$
MeV are chosen. The ratio between $\sigma_{np}^*$ and
$\sigma_{pp(nn)}^*$ for $Y_p=0.5$ is shown in the upper-right
plot.} \label{fig2}
\end{figure}

The main contributions to the in-medium ECS come from the
$\sigma-\sigma$, $\omega-\omega$ and the crossing term
$\sigma-\omega$. Fig. \ref{fig2} (a) shows the contributions of
$\sigma-\sigma$, $\omega-\omega$ and $\sigma-\omega$ terms to
$\sigma_{pp(nn)}$ and $\sigma_{np}$ at $T=0$ MeV, $E_{K}=10$ MeV
for $Y_p=0.5$. The ratio between $\sigma_{np}$ and
$\sigma_{pp(nn)}$ is also shown in the upper-right plot. From this
figure one can find both $\sigma_{pp(nn)}$ and $\sigma_{np}$ have
similar shape of density dependence and the
$\sigma_{np}$/$\sigma_{pp(nn)}$ ratio is almost density
independent.

\begin{figure}
\includegraphics[angle=0,width=0.5\textwidth]{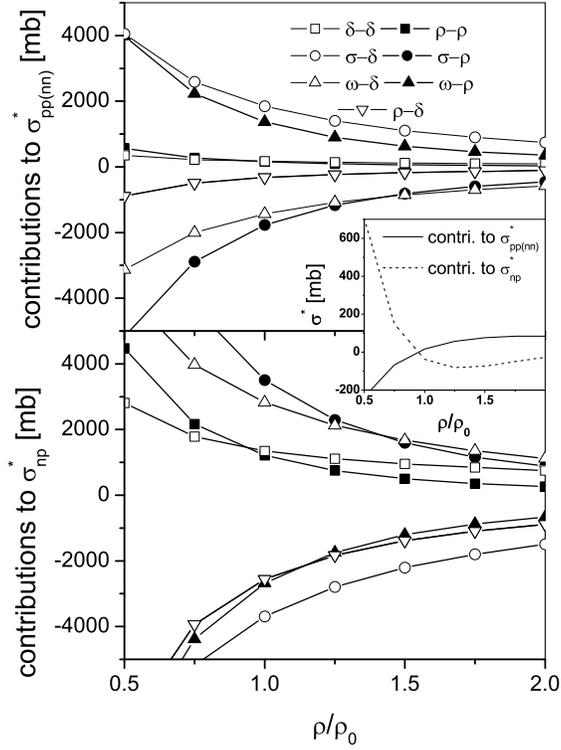}
 \caption{
The contributions from the $\delta$, $\rho$ meson related terms to
$\sigma_{pp(nn)}$ (in upper plot) and $\sigma_{np}$ (in lower
plot) at $T=0$ MeV and $E_K=10$ MeV for $Y_{p}=0.5$. The
central-right small plot shows the sum of the contributions from
the $\delta$ and $\rho$ meson related terms.} \label{fig3}
\end{figure}

The contributions from the $\delta$ and $\rho$ mesons and related
crossing terms are shown in Fig. \ref{fig3}. In order to explore
how $\delta$ and $\rho$ meson field affect the in-medium ECS, we
show the individual contributions from each term. The sub-figure
shows the sum of the $7$ terms contributing to $\sigma_{pp(nn)}$
and $\sigma_{np}$. This figure shows rich information about the
contributions of each term to the in-medium ECS as well as the
cancellation effect among the $7$ terms. Firstly, the magnitude of
all individual contributions to the in-medium ECS decreases with
density. This density dependence results from both the density
dependence of the meson-nucleon couplings and the effective masses
of protons and neutrons. Among the $7$ terms, the crossing terms
of the $\sigma$ meson to $\delta$ and $\rho$ mesons provide the
most important contributions and the crossing terms of $\omega$ to
$\delta$ and $\rho$ mesons are the next important terms. Further,
the cancellation among the contributions from different terms are
very strong so the final results are the delicate balance of all
contributed terms.

In Fig. \ref{fig2} (b) we show the density dependence of
$\sigma_{pp}^*$, $\sigma_{nn}^*$, and $\sigma_{np}^*$ at $T=0$ MeV
for $Y_{p}=0.3$ and $Y_{p}=0.5$. The $\sigma_{pp(nn)}^*$ and
$\sigma_{np}^*$ for $Y_{p}=0.5$ are shown with thick line types
while those for $Y_{p}=0.3$ are illustrated with light lines. One
can see from the figure that $\sigma_{pp}^*$, $\sigma_{nn}^*$, and
$\sigma_{np}^*$ decrease with the increase of densities. This
behavior is consistent with other theoretical calculations, for
instance, Refs. \cite{LiGQ93,Li96,LiGQ2} and the phenomenological
expression of $\sigma^*=[1-\alpha(\rho/\rho_0)]\sigma^{free}$. The
most impressive thing shown in the figure is that the density
dependence of $\sigma_{np}^*$ and $\sigma_{pp(nn)}^*$ is very
different: At low densities, the $\sigma_{np}^*$ is about $3-4$
times larger than $\sigma_{pp(nn)}^*$ and then decreases quickly
with density and finally the $\sigma_{np}^*/\sigma_{pp(nn)}^*$
ratio approaches to $\sim 1$ as density reaching about the normal
density. The $\sigma_{np}^*/\sigma_{pp(nn)}^*$ ratio is shown in
the small top-right plot.  It means that at dense nuclear matter
the isospin effect on in-medium ECS almost washes out. For
$Y_p=0.3$ case, $\sigma_{pp}^*$ and $\sigma_{nn}^*$ differ. The
value of $\sigma_{pp}^*$ is larger than that of $\sigma_{nn}^*$
because $M_p^*>M^*>M_n^*$ (see Fig. \ref{fig1} (a)). The splitting
of $\sigma_{pp}^*$ and $\sigma_{nn}^*$ increases with density and
the degree of isospin asymmetry of the medium following the
splitting of the proton and neutron effective mass.

In Ref. \cite{Li00} the in-medium $\sigma_{np}^*$ and
$\sigma_{pp(nn)}^*$ are calculated based on QHD-II model with the
medium correction of the effective mass of $\rho$ meson taken into
account. We found in \cite{Li00} that $\sigma_{pp(nn)}^*$ depends
on density obviously but $\sigma_{np}^*$ depends on density
weakly. The different results can be attributed to the inclusion
of the $\delta$ mesons. In the mean field level, the contributions
of the $\delta$ and $\rho$ mesons compensate each other in central
potential, producing an effective isovector potential that is
comparable in strength to the one obtained in calculations that
include only the $\rho$ meson. While for the calculations of the
in-medium cross sections, which is beyond the mean field, the
direct compensation of the contributions from $\delta$ and $\rho$
mesons is lost. So we may not expect that the comparable results
for the in-medium cross sections should be obtained for the cases
of inclusion and not inclusion of the $\delta$ meson. The
different results for the density dependence of
$\sigma_{pp(nn)}^*$ and $\sigma_{np}^*$ obtained in two different
models may provide us with a more strict check for the model.

\begin{figure}
\includegraphics[angle=0,width=0.8\textwidth]{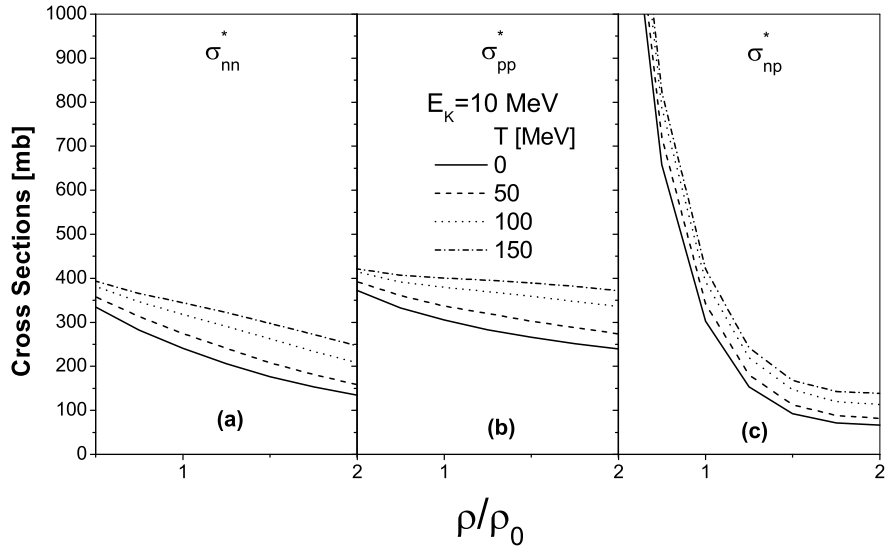}
 \caption{
 $\sigma_{pp(nn)}^*$ and $\sigma_{np}^*$ at $T=0$, $50$, $100$, and $150$ MeV as
 function of density for $Y_{p}=0.3$ and $E_K=10$ MeV.}
\label{fig4}
\end{figure}

Now let us investigate the temperature effect on
$\sigma_{pp(nn)}^*$ and $\sigma_{np}^*$. Fig. \ref{fig4}  shows
$\sigma_{nn}^*$ (in (a)), $\sigma_{pp}^*$ (in (b)), and
$\sigma_{np}^*$ (in (c)) at $T=0$, $50$, $100$, and $150$ MeV. In
Fig. \ref{fig4} one sees that the in-medium ECS increases with
temperature and the influence of temperature on in-medium ECS
increases with density. We should point out that this temperature
effect is model dependent. With the model similar to that used in
Ref. \cite{Li00} we find the temperature effect on the medium
correction of nuclear mass and then on the ECS is weaker than the
present results. One of the most important information contained
in Fig. \ref{fig4} is that the influence of temperature on
in-medium cross sections is much weaker than that of density. The
cause of this difference can be attributed to the different
behavior of the density and temperature dependence of the nucleon
effective mass. This tells us that one should consider the effect
of medium density on the in-medium cross section more seriously
than that of medium temperature.

In summary, in this work we have studied the density and
temperature effects on in-medium elastic cross sections based on
the density dependent relativistic hadron field theory in which
the $\delta$ meson is included. Our results show that at low
densities the $\sigma_{np}^*$ is about $3-4$ times larger than
$\sigma_{pp(nn)}^*$ and then decreases quickly with density and
finally the $\sigma_{np}^*/\sigma_{pp(nn)}^*$ ratio approaches to
$\sim 1$ as densities reaching and exceeding the normal density.
It means that the isospin effect on in-medium elastic cross
section almost washes out at high densities. The analysis about
the individual contributions from each term contributing to
$\sigma_{pp(nn)}^*$ and $\sigma_{pn}^*$ indicates that the isospin
effect on the density dependence of the in-medium elastic cross
sections is dominantly caused by the delicate balance of the
isovector $\delta$ and $\rho$ mesons. For isospin asymmetric
medium, the $\sigma_{nn}^*$ and $\sigma_{pp}^*$ differ following
the splitting of the proton and neutron mass due to the inclusion
of the $\delta$ meson. Our calculation results show that the
in-medium ECS increases with the temperature and the temperature
effect of nuclear medium on $\sigma_{np}^*$ and
$\sigma_{pp(nn)}^*$ is weaker compared with the density effect.
The temperature effect enhances as density increases so the
influence of temperature to in-medium ECS may have to be taken
into account at higher nuclear density.

The work is supported by the National Natural Science Foundation
of China under Nos. 10175093 and 10235030, Science Foundation of
Chinese Nuclear Industry and Major State Basic Research
Development Program under Contract No. G20000774, the Knowledge
Innovation Project of the Chinese Academy of Sciences under Grant
No. KJCX2-SW-N02, and the CASK.C. Wong Post-doctors Research Award
Fund.

\end{document}